\documentclass[twocolumn]{aastex61}
\usepackage{graphicx}
\graphicspath{{./plot/}}
\usepackage{epstopdf}
\usepackage{hyperref}
\usepackage{xcolor}
\usepackage{wrapfig}
\usepackage[utf8x]{inputenc}
\usepackage{rotating}
\usepackage{natbib}
\usepackage{mathtools}
\usepackage[varg]{txfonts}

\begin{document}

\title{COMPLETE IRAC MAPPING OF THE CFHTLS-DEEP, MUSYC  and NMBS-II FIELDS}

\author{Marianna Annunziatella}
\affil{Physics Department, Tufts University, 574 Boston Ave., Medford, 02155, MA}

\author{Danilo Marchesini}
\affil{Physics Department, Tufts University, 574 Boston Ave., Medford, 02155, MA}

\author{Mauro Stefanon}
\affil{Leiden Observatory, Leiden University, PO Box 9513, NL-2300 RA, Leiden, The Netherlands}

\author{Adam Muzzin}
\affil{Department of Physics and Astronomy, York University, 4700 Keele St., Toronto, Ontario, MJ3 1P3, Canada}

\author{Daniel Lange-Vagle}
\affil{Physics Department, Tufts University, 574 Boston Ave., Medford, 02155, MA}

\author{Ryan Cybulski}
\affil{Physics Department, Tufts University, 574 Boston Ave., Medford, 02155, MA}

\author{Ivo Labbe'}
\affil{Centre for Astrophysics and SuperComputing, Swinburne, University of Technology, Hawthorn, Victoria, 3122, Australia}

\author{Erin Kado-Fong}
\affil{Department of Astrophysical Sciences, Princeton University, Princeton, NJ 08544, USA}

%%% in alphabetical order
\author{Rachel Bezanson}
\affil{Department of Physics and Astronomy and PITT PACC, University of Pittsburgh, Pittsburgh, PA 15260, USA}

\author{Gabriel Brammer}
\affil{Space Telescope Science Institute, 3700 San Martin Drive, Baltimore, MD 21218, USA}

%\author{Gabriella De Lucia}
%\affil{INAF-Osservatorio Astronomico di Trieste, via G. B. Tiepolo 11, 34131, Trieste, Italy}

%\author{Mark Dickinson}
%\affil{National Optical Astronomy Observatory, 950 North Cherry A venue, Tucson, AZ 85719, USA}

%\author{Marijn Franx}
%\affil{Leiden Observatory, Leiden University, PO Box 9513, NL-2300 RA, Leiden, The Netherlands}

\author{David Herrera}
\affil{National Optical Astronomy Observatory, 950 North Cherry Avenue, Tucson, AZ 85719, USA}

\author{Britt Lundgren}
\affil{Department of Physics, University of North Carolina Asheville, One University Heights, Asheville, NC 28804, USA}

\author{Z. Cemile Marsan}
\affil{Department of Physics and Astronomy, York University, 4700 Keele St., Toronto, Ontario, MJ3 1P3, Canada}

\author{Mario Nonino}
\affil{INAF-Osservatorio Astronomico di Trieste, via G. B. Tiepolo 11, 34131, Trieste, Italy}

\author{Gregory Rudnick}
\affil{Department of Physics and Astronomy, University of Kansas, 1251 Wescoe Hall Dr., Room 1082, Lawrence, KS, USA}

\author{Paolo Saracco}
\affil{INAF-Osservatorio Astronomico di Brera, Via Brera 28, I-20121 Milano, Italy}

\author{Tal Tomer}
\affil{ Department of Astronomy and Astrophysics, University of California, Santa Cruz, CA, USA}

\author{Frank Valdes}
\affil{National Optical Astronomy Observatory, 950 North Cherry A venue, Tucson, AZ 85719, USA}

\author{Remco F. J. van der Burg}
\affil{European Southern Observatory, Karl-Schwarzschild-Str. 2, 85748, Garching, Germany}

\author{Pieter van Dokkum}
\affil{Astronomy Department, Yale University, New Haven, CT 06511, USA}

\author{David Wake}
\affil{Department of Physics, University of North Carolina Asheville, One University Heights, Asheville, NC 28804, USA}

\author{Katherine E. Whitaker}
\affil{Department of Physics, University of Connecticut, 2152 Hillside Road, Unit 3046, Storrs, CT 06269, USA}
\affil{Cosmic Dawn Center (DAWN), Niels Bohr Institute, University of Copenhagen / DTU-Space, Technical University of Denmark}

\begin{abstract}
The IRAC mapping of the NMBS-II fields program is an imaging survey at 3.6 and 4.5$\mu$m with the Spitzer Infrared Array Camera (IRAC). 
%A correct estimate of $z_{phot}$ and $M_{\star}$ is crucial to identify Ultra Massive galaxies (UMGs) which represent the biggest problem in our understanding of how galaxy forms. 
The observations cover three Canada-France-Hawaii Telescope Legacy Survey Deep (CFHTLS-D) fields, including  one also imaged by AEGIS, and two MUSYC fields. These are then combined with archival data from all previous programs into deep mosaics. The resulting imaging covers a combined area of about 3 $deg^2$, with at least $\sim$2 hr integration time for each field. In this work, we present our data reduction techniques and document the resulting coverage maps  at 3.6 and 4.5$\mu$m. All of the images are W-registered to the reference image, which is either the z-band stack image of the 25\% best seeing images from the CFHTLS-D for CFHTLS-D1, CFHTLS-D3, and CFHTLS-D4, or the K-band images obtained at the Blanco 4-m telescope at CTIO for MUSYC1030 and MUSYC1255.
We make all images and coverage maps described herein publicly available via the Spitzer Science Center.
%These images will allow us to classify the population of galaxies at z$>$0.8 into quiescent and star-forming galaxies using the UVJ diagram, and to obtain more accurate estimate of $z_{phot}$ and $M_{\star}$. 
\end{abstract}


\graphicspath{{./}}

% % % % % aggiungere a e dec dei campi. 
\section{Introduction}

One of the major unsolved issues in astrophysics is how galaxy buildup proceeded in the early Universe. Evolutionary models predict that different mechanisms can play a fundamental role for galaxy growth, such as galaxy mergers \citep[e.g.][]{hopkins2006, somerville2008} or cold gas accretion within gas-rich proto-disks \citep[e.g.][]{dekel2009}. Whatever the physical mechanisms dominating galaxy evolution might be, they must reproduce the observed distribution of baryons at high redshift, and connect it to the subsequent evolution of galaxies within dark matter halos. In this context, it becomes crucial to have accurate measurements of galaxy number densities and stellar-mass evolution over most of the cosmic history ($\mathrm{z < 3 - 4}$). So far, important discrepancies still exist between predictions from theoretical models and observations of the number density of the most massive galaxies ($\mathrm{M_{\star}>3×10^{11} M_{\odot}}$) at high redshifts \citep[eg.,][]{fontanot2009, marchesini2009,marchesini2010, ilbert2013, hirschmann2016}. In fact, theoretical models still under-predict the observed number density of these galaxies, unless the model-predicted stellar mass functions (SMFs) are convolved with large errors in  $\mathrm{M_{\star}}$ \citep[e.g., $\sim \, 0.25-0.3$ dex at z=2-3,][]{marchesini2009,marchesini2010,ilbert2013, muzzin2013b}, arguably much larger than the typical random errors on $\mathrm{M_{\star}}$ for massive (hence bright) galaxies.
This disagreement at the very high-mass end has been recently confirmed by recent surveys, e.g. the NEWFIRM (NOAO Extremely Wide-Field Infrared Imager ) Medium-Band Survey (NMBS) \citep{whitaker2011} and the UltraVISTA survey \citep{muzzin2013a, hirschmann2016}. 
NMBS is a moderately deep survey covering 0.45 $\mathrm{deg^2}$ over two $\mathrm{30\arcmin \times 30 \arcmin}$ fields within the Cosmic Evolution Survey \citep[COSMOS;][]{scoville2007} and the All-Wavelength Extended Groth Strip International Survey \citep[AEGIS;][]{davis2007}, chosen to take advantage of the wealth of publicly available ancillary data over a broad wavelength range, from the X-ray to the radio.
 Thanks to the medium-band near-infrared (NIR) filters,  NMBS provides precise $\mathrm{z_{phot}}$ ($\mathrm{\Delta z/(1 + z)\approx 0.02}$) and well-sampled spectral energy distributions (SEDs) for thousands of K-selected galaxies at $z>1.5$. \\
One of the most remarkable results from NMBS is the clear detection of the galaxy bi-modality in the galaxy population (quiescent vs. star-forming) out to $\mathrm{z \sim 2.5}$ \citep{brammer2009}. 
The NMBS showed that nearly all galaxies with $\mathrm{log(M/M_{star})\, > \, 10.5 }$ have red rest-frame U−V colors up to z $\mathrm{=}$ 2.  Despite similar U−V colors, the NMBS data showed that these galaxies have strikingly different properties, with approximately half being red due to a lack of recent star formation (quiescent), the other half being red due to vigorous dusty star formation. This result showed the limitations of using a single color to identify quiescent galaxies at $\mathrm{z\,>\,1}$. Instead, star-forming and quiescent galaxies can be robustly separated out to $\mathrm{z \sim\, 3}$ with a two-color criterion using the rest-frame U$-$V versus V$-$J diagram (UVJ diagram, hereafter).
%Provided accurate enough zphot and well-sampled SEDs, the U−V and V −J colors are robustly measured directly from the observed fluxes, whereas SFR or sSFR are among the most uncertain stellar population properties to be estimated, even with spectroscopic redshifts (e.g., [56]). 
However, for high redshift galaxies, $\mathrm{3.6\, and\,  4.5\, \mu m}$ imaging with IRAC \citep[][]{fazio2004} aboard the Spitzer Space Telescope \citep{werner2004} are absolutely critical for the UVJ diagnostic, as the rest-frame V $-$ J color cannot be directly measured without IRAC for galaxies at $\mathrm{0.8\,< \, z\, <\, 3}$.  
It is only thanks to IRAC in combination with the very precise photometric redshift ($\mathrm{z_{phot}}$) from NMBS that the existence of the bi-modality and the existence of quiescent galaxies out to z$\sim$2.5 (and beyond) have been shown definitively \citep{whitaker2011}. The lack of IRAC photometry prevents robust classification into quiescent and star-forming galaxies at $z>0.8$.\\
While our knowledge of massive galaxies has grown significantly in the past years due to NMBS and other many NIR surveys, little is known about the evolution of the most massive galaxies  (UMGs) with $\mathrm{M_{\star} >3\times10^{11}M_{\odot}}$.  This is simply because their space density is extremely low ($\sim$30× lower than galaxies with $\mathrm{M_{\star} =10^{11}M_{\odot}}$). Even the largest NIR surveys have lacked the volume to detect them in significant numbers. Despite their rareness, these galaxies arguably represent the biggest problem in our understanding of how galaxies form. Abundance matching models suggest that  ultra-massive galaxies, those with $\mathrm{M_{\star} =3\times 10^{11}M_{\odot}}$  should reside in dark matter haloes of a few $\mathrm{\times 10^{14}M_{\odot}}$ at all redshifts, i.e., they are the progenitors of local Brightest Cluster Galaxies. While observationally the most massive galaxies appear to be the first to become quiescent, it is completely unknown why this is the case and what drives the quenching of the star formation. \\
In 2010-2012, NMBS-II (PI: Van Dokkum),  a 44-night NOAO survey program with NEWFIRM was conducted with the primary goal of surveying a large enough volume to identify a statistically robust sample of UMGs out to $z=3.5$. NMBS-II is shallower but $11\times$ wider than NMBS, and it is unique for its combination of NIR medium-band imaging, well-sampled SEDs, and large area. NMBS-II covers a total of $\sim$ 5.4 $\mathrm{deg^2}$ over 7 fields, namely CFHTLS-D1, CFHTLS-D2, (COSMOS), CFHTLS-D3, CFHTLS-D4, and 3 MUSYC (Multiwavelength Survey by Yale-Chile) fields, namely MUSYC1030, MUSYC1255 and MUSYC-ECDF. 
Observations of the NMBS-II are now completed, and multi-wavelength K-selected catalogs have been constructed (including deep optical ugriz data from the CFHTLS and U-V data from Galex). The NMBS-II J1J2J3H1H2K data, combined with deep optical data from the CFHTLS and IRAC imaging over the full survey area, promise to provide a mass-complete sample of $\sim$ 300 UMGs ($\mathrm{log(M_{\star}/M_{\odot}) > 11.4}$, of which $\sim$50 with $\mathrm{log(M_{\star}/M_{\odot}) > 11.6}$) with high-confidence $\mathrm{z_{phot}}$ at $1.5<z<3$. This sample, spanning the full range of star-formation activity, will provide the first comprehensive view of the population of UMGs at $z>1.5$. This will be the largest sample of UMGs with accurate $\mathrm{z_{phot}}$ ever constructed. 
NMBS-II is truly unique as no other current survey has its combination of surveyed area and NIR medium-band coverage, preventing the delivery of sufficiently precise $\mathrm{z_{phot}}$ over a wide enough area to assemble similarly large samples of UMGs at $z>1.5$.\\
However, Spitzer-IRAC coverage from previous programs of NMBS-II fields is available only for $\sim$60\% of the surveyed  area, hence missing for $\sim 2.2 \mathrm{deg^2}$.  IRAC data, besides being critical to classify the population of galaxies at z$>$0.8 into quiescent and star-forming galaxies using the UVJ diagram, are crucial for more accurate estimates of $z_{phot}$ and $M_{\star}$, as well as for robust determinations of SFRs, as they significantly help breaking the age-dust degeneracy in the modeling of the SEDs.   \\
With this paper, we present and publicly release the scientific mosaics, coverage maps, and 3$\sigma$ depth maps  at 3.6$\mu$m and 4.5$\mu$m for five of the NMBS-II fields. The mosaics combine both archival data and new data observed in Cycle 10. 
The paper is organized as follows. Section \ref{s:obs} describes all the programs that have been used to create single, contiguous deep images in the 3.6 and 4.5$\mu m$ bands. In Section \ref{s:reduction}, we describe how the mosaics and coverage maps were produced, and Section~\ref{s:prop} presents the characterization of the mosaics, including image quality and depth.
All coordinates refer to the J2000 system. All magnitudes are in the AB system.

\section{ Observations}\label{s:obs}
In this section,  we present the NMBS-II fields and the individual Spitzer-IRAC programs that have been combined to produce deep IRAC mosaics over the NMBS-II, completing coverage of the full survey. Although the NMBS-II survey is comprised of seven fields, we produce IRAC mosaics only for five fields, because comprehensive IRAC mosaics in COSMOS and ECDFS are already publicly available from the Spitzer Science Center as part of previous IRAC surveys.
% in COSMOS (\citealt[i.e., S-COSMOS,][]{sanders2007},SEDS, \citealt{ashby2013}; S-CANDELS, \citealt{ashby2015}; SPLASH,  \citealt{steinhardt2014}; and Spitzer Matching survey of the UltraVISTA ultra-deep Stripes \citep[SMUVS;][]{ashby2018}) and in CDFS (i.e., GOODS-IRAC, \citealt{dickinson2003}; and SIMPLE, \citealt{damen2011}). 

\subsection{NMBS-II Fields}\label{ss:nmbs-ii fields}
In this sub-section, we present the seven NMBS-II fields, including the ancillary data publicly available.

\begin{itemize}
\item CFHTLS-D1 (XMM-LSS) is a $\mathrm{1 \, deg^2}$ field centered on RA= 02:26:00.00 (hh:mm:ss) and DEC= -04:30:00.00 (dd:mm:ss). This field has ugriz imaging from the Canada-France-Hawaii Telescope Legacy Survey (CFHTLS).
% It was also observed as part of the XMM-LSS, GALEX Deep imaging survey, and has spectroscopic follow-up by VIMOS.
 This field has NIR  images in YJHK taken  as part of the Visible and Infrared Survey Telescope for
Astronomy (VISTA) Deep Extragalactic Observations (VIDEO) survey \citep{jarvis2013}. Other NIR data are available through the WIRCam Infrared Deep Survey \citep[WIRDS;][]{bielby2012}. Imaging of the CFHTLS-D1 field at 250, 350, and 500 $\mu$m was
provided by the Spectral and Photometric Imaging REceiver \citep[SPIRE;][]{griffin2010} aboard the Herschel Space Observatory \citep{pilbratt2010}, taken as part of the Herschel Multi-tiered Extragalactic Survey \citep[HerMES;][]{oliver2012} X-ray observations of the CFHTLS-D1 field were taken from the X-ray Multi-mirror Mission space telescope \citep[XMM-Newton;][]{jansen2001} as part of the XMM Medium Deep Survey \citep[XMDS;][]{chiappetti2005}. Spectroscopic data for this field have been acquired with the VIsible Multi-Object Spectrograph (VIMOS) as part of the VIMOS-VLT deep survey (VVDS).
As a part of NMBS-II, this field was imaged in medium-bandwidth filters ($J_1J_2J_3H_1H_2$) over the wavelength range $\mathrm{1–1.8\mu m}$ and in the K-band.
CFHTLS-D1 has a  100\% data coverage from the Spitzer Wide-area InfraRed Extragalactic Survey \citep[SWIRE,][]{lonsdale2003}. SWIRE is a wide-area, high galactic latitude imaging survey which covered  roughly 50 $\mathrm{deg^2}$ across a number of fields with the MIPS far-infrared camera and the with the IRAC mid-infrared camera. 

\item CFHTLS-D2  is a $\mathrm{2 \, deg^2}$ field centered on  RA= 10:00:28.60 and DEC= 02:12:21.00. This field is part of the Cosmic Evolution Survey survey. \citep[COSMOS;][]{scoville2007}. Within COSMOS, $\mathrm{1\, deg^2}$ has is also been imaged by CFHT as part of the Canada-France-Hawaii Telescope Legacy Survey Deep (CFHTLS-Deep). This field has X-ray imaging \citep[Chandra and XMM,][respectively]{elvis2009,finoguenov2007} from COSMOS, HST imaging from the original COSMOS Hubble Space Telescope Treasury project, and from the Cosmic Assembly Near-infrared Deep Extragalactic Legacy Survey \citep[CANDELS;][]{koekemoer2011}, deep optical imaging taken with Hyper Suprime-Cam \citep[HSC;][]{tanaka2017}, and deep NIR imaging as part of UltraVISTA \citep{mckracken2013} survey.  As a part of NMBS-II, this field was imaged in medium-bandwidth filters ($\mathrm{J_1 J_2 J_3 H_1 H_2}$),  and in the K-band.
Narrow-band NIR images of the COSMOS-HST field we also taken with Suprime-Cam on the Subaru Telescope within the Subaru COSMOS 20 project \citep{taniguchi2015}. COSMOS has extensive spectroscopic data with over 97,000 spectra targeting over 68,000 unique objects.  Major surveys include zCOSMOS \citep{lilly2007}, and the VIMOS Ultra Deep Survey \citep[VUDS;][]{lefevre2015}. HST spectroscopy is also available from the 3D-HST program \citep{vandokkum2011}. CFHTLS-D2 has 100\% IRAC data coverage coming from multiple surveys, e.g. S-COSMOS \citep{sanders2007}, the Spitzer Extended Deep Survey  \citep[SEDS;]{ashby2013}, S-CANDELS, \citep{ashby2015}, Star Formation at 4 $<$ z $<$ 6 from the Spitzer Large Area Survey with Hyper-Suprime-Cam \citep[SPLASH;][]{steinhardt2014}, and Spitzer Matching survey of the UltraVISTA ultra-deep Stripes \citep[SMUVS;][]{ashby2018}.

\item CFHTLS-D3  is a $\mathrm{1\, deg^2}$ field centered on RA= 14:19:27.00 and DEC= +52:40:56.00. This field was imaged by CFHT both as part of the AEGIS survey  and CFHTLSD. X-ray images of this field were taken with Chandra as part of the  AEGIS-X \citep{laird2009}.  Deep HST images of the EGS were obtained with the Advanced Camera for Surveys (ACS) as part of GO program 10134 (PI: M. Davis). As a part of NMBS-II, this field was imaged in medium-bandwidth filters ($\mathrm{J_1 J_2 J_3 H_1 H_2}$),  and in the K-band. CFHTLS-D3 previously had 20\% IRAC data coverage from  S-CANDELS, \citep{ashby2015}. 

\item CFHTLS-D4  is a $\mathrm{1 \, deg^2}$ field centered on RA= 22:15:31.00  and DEC= -17:43:56.00. This field was imaged by CFHT as part of CFHTLS. As a part of NMBS-II, this field was imaged in medium-bandwidth filters ($\mathrm{J_1 J_2 J_3 H_1 H_2}$),  and in the K-band.
CFHTLS-D4 had no IRAC coverage previous to this paper.

\item MUSYC Extended CDFS (MUSYC-ECDFS) is a 0.5 $\mathrm{deg^2}$ field centered on RA= 03:32:29.0 and DEC= -27:48:47.00. This is one of the  fields of the Great Observatories Origins Deep Survey \citep[GOODS;][]{dickinson2001}. Within GOODS, the ACS camera on
HST acquired imaging of the ECDFS field. Ground-based optical imaging of the ECDFS with the Wide Field Imager (WFI) on the ESO/MPG 2.2m telescope was obtained as part of the COMBO-17 \citep{wolf2004}. NIR images of this field were also taken with the VLT/ISAAC.
This field has also NIR  images in YJHK taken  as part of VIDEO survey \citep{jarvis2013}. As a part of NMBS-II this field was imaged in medium-bandwidth filters ($\mathrm{J_1 J_2 J_3 H_1 H_2}$),  and in the K-band.
MUSYC-ECDFS has 100\% data coverage from the Spitzer IRAC/MUSYC Public Legacy Survey in the Extended Chandra Deep Field South \citep[SIMPLE,][]{damen2011}. SIMPLE consists of deep IRAC observations (several hours per pointing) covering the 0.5 x 0.5 $\mathrm{deg^2}$ area surrounding the GOODS CDFS field.  
\item MUSYC1030 is a 0.5 $\mathrm{deg^2}$ field centered around RA= 10:30:27.10 and DEC= 05:24:55.02. The center of this field is the position of the SDSS quasar SDSS J1030+0524S \citep{becker2001}. This field was imaged by different telescopes within several surveys. Optical images of this field were taken within the MUSYC survey \citep{gawiser2006a}. Complimentary K-band images were taken using the Infrared Side Port Imager (ISPI) on the 4m Blanco Telescope at the Cerro Tololo Inter-American Observatory (CTIO) in the MUSYC wide NIR survey. As a part of NMBS-II, this field was imaged in medium-bandwidth filters ($\mathrm{J_1 J_2 J_3 H_1 H_2}$),  and in the K-band. MUSYC1030 has 10\%  IRAC coverage.

\item MUSYC1255 is a 0.5 $\mathrm{deg^2}$ centered at RA= 12:55:40.0 and DEC= 01:07:00.00 and also has optical and NIR  imaging as part of the 
MUSYC and MUSYC-NIR surveys. As a part of NMBS-II, this field was imaged in medium-bandwidth filters ($\mathrm{J_1 J_2 J_3 H_1 H_2}$),  and in the K-band. MUSYC1255 has 10\%  IRAC coverage from the Spitzer Space Telescope Cycle-3 program GO-30873 (PI: Labbe ). 
\end{itemize}

\subsection{New IRAC imaging in NMBS-II}\label{ss: new_irac}
To fully exploit the NMBS-II,  in Cycle 10, we were awarded 22 hours on Spitzer to complete the IRAC coverage of the NMBS-II fields in $\mathrm{3.6\, \mu m}$ (3.6$\mu$m) and $\mathrm{4.5\, \mu m}$ (4.5$\mu$m) (ID:10084; PI: D. Marchesini). The fields imaged in this program were: CFHTLS-D3, CFHTLS-D4,MUSYC1030, andMUSYC1255. The total area mapped in $\mathrm{3.6\, \mu m}$ and $\mathrm{4.5\, \mu m}$ in Cycle 10  is $\sim$ 2.2 $\mathrm{deg^2}$, for a total of 342 individual pointings.\\
In this paper, we present scientific mosaics and coverage maps for CFHTLS-D1,  CFHTLS-D3, CFHTLS-D4,MUSYC1030, andMUSYC1255.

\section{Data reduction}\label{s:reduction}

We downloaded all archival data from the Spitzer Heritage Archive (SHA) and combined them with our data sets. We reduced all data in a consistent manner, coadding them into one mosaic per field. We followed the procedure described in \citep{labbe2015}.
The individual exposures are organized into self-contained segments known as Astronomical Observing Requests (AORs). Over the five fields, there are 755 AORs for each filter.  Each AOR consists of basic calibrated data (BCD). BCDs are exposure-level data after having passed through the pipelines. Instrumental signatures have been mostly removed, and the BCDs have been absolutely calibrated into physical units (i.e., $MJy/sr =  10^{-17} erg s^{-1} cm^{-2} Hz^{-1} sr^{-1}$).  The BCDs were improved by applying a series of procedures in order to remove cosmic rays and remaining instrumental artifacts including column pulldown,  muxbleed, maxstripe, and the ``first-frame effect'' \citep{hora2004}.  Single background-subtracted frames were then combined into a mosaic image large enough to hold all input frames (using WCS astrometry  to align the images). In this step,  we masked bad pixels and applied a distortion correction in WCS. This image, in combination with the reference image, was then used to refine the pointing of the individual mosaics. The individual pointing-refined frames were registered to and projected on the reference image. \\
The reference images for the CFHTLSD-1, CFHTLSD-3, CFHTLSD-4 fields are the $z-$band stack from CFHTLSDeep taken from the CFHTLS  T0007 release. These stacks are the combinations of the 25\% best seeing images (seeing $< 0.95 \arcsec$) of each fields in the  $z-$band.  Each stack comprises all images located inside a radius of $3 \arcmin$ with respect to a tile center position,  weighted accordingly, and combined using SWarp, with a Lanczos3 interpolation kernel. The total exposure times for each of the reference images is 14.20hr, 12.51hr and 12.00 hr, while the seeing is 0.559$\arcsec$, 0.537 $\arcsec$, and 0.551$\arcsec$, for CFHTLS-D1, CHFTLS-D3, and CFHTLS-D4 respectively. The pixel scale is 0.186 $\arcsec / pixel$. These images are available on the Terapix database (\url{http://terapix.iap.fr/cplt/T0007/table_syn_T0007.html}).\\
The final IRAC mosaics and coverage maps are then re-sampled to a higher pixel scale value which is listed in Table~\ref{tab-appiracdata1}. 
The reference images for MUSYC1030 and MUSYC1255 are taken with  the Infrared Sideport Imager (ISPI)  camera mounted at the prime focus of the Blanco 4-m telescope at CTIO in the K-band filter. The original detector pixel scale is 0.305$\arcsec$/pixel. However both K-band images have been previously processed so that the pixel scale is changed slightly to 0.267 and 0.269 $\arcsec$/pixel, respectively for MUSYC1030 and MUSYC1255.
Table~\ref{tab-appiracdata1} summarizes the characteristics of the IRAC data over the five observed fields, such as the total exposure time, the number of AORs that contributed to the mosaic, and the pixel scale at $\mathrm{3.5 \, \mu m}$ and $\mathrm{4.5 \, \mu m}$. For comparison, we also add, where available, the characteristics of the other two fields.\\

\begin{figure*}[ht]
	\centering
	\includegraphics[scale=0.5]{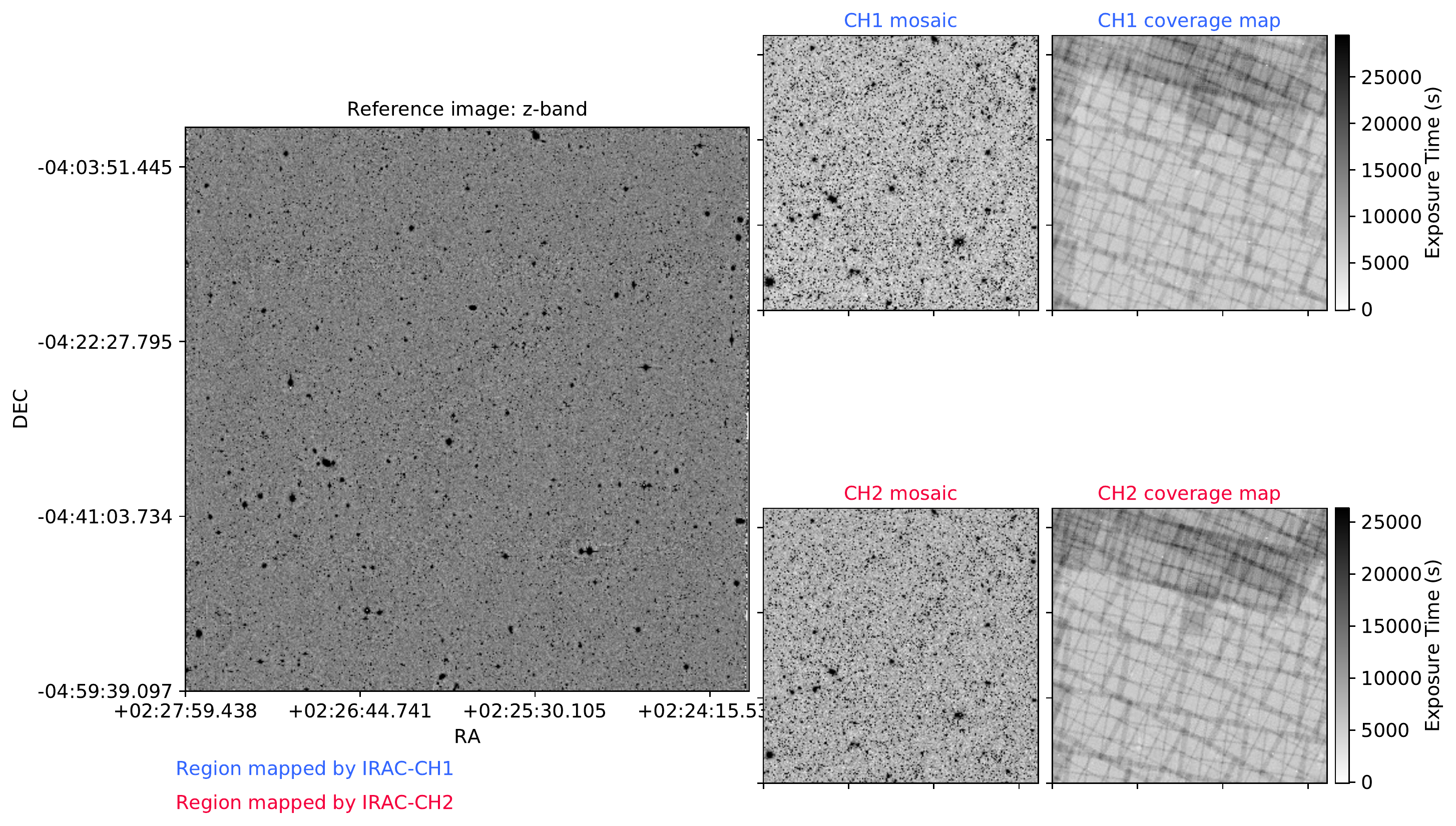}
	\caption{ CFHTLS-D1: reference band mosaic, 3.6$\mu m$ and 4.5$\mu m$ mosaics and coverage maps. The coverage maps show where the individual pointings of IRAC overlap when they are combined to form the mosaic. Pixels with darker colors have more coverage.}
	\label{f:cfhtlsd1}
\end{figure*}

\begin{figure*}[ht]
	\centering
	\includegraphics[scale=0.5]{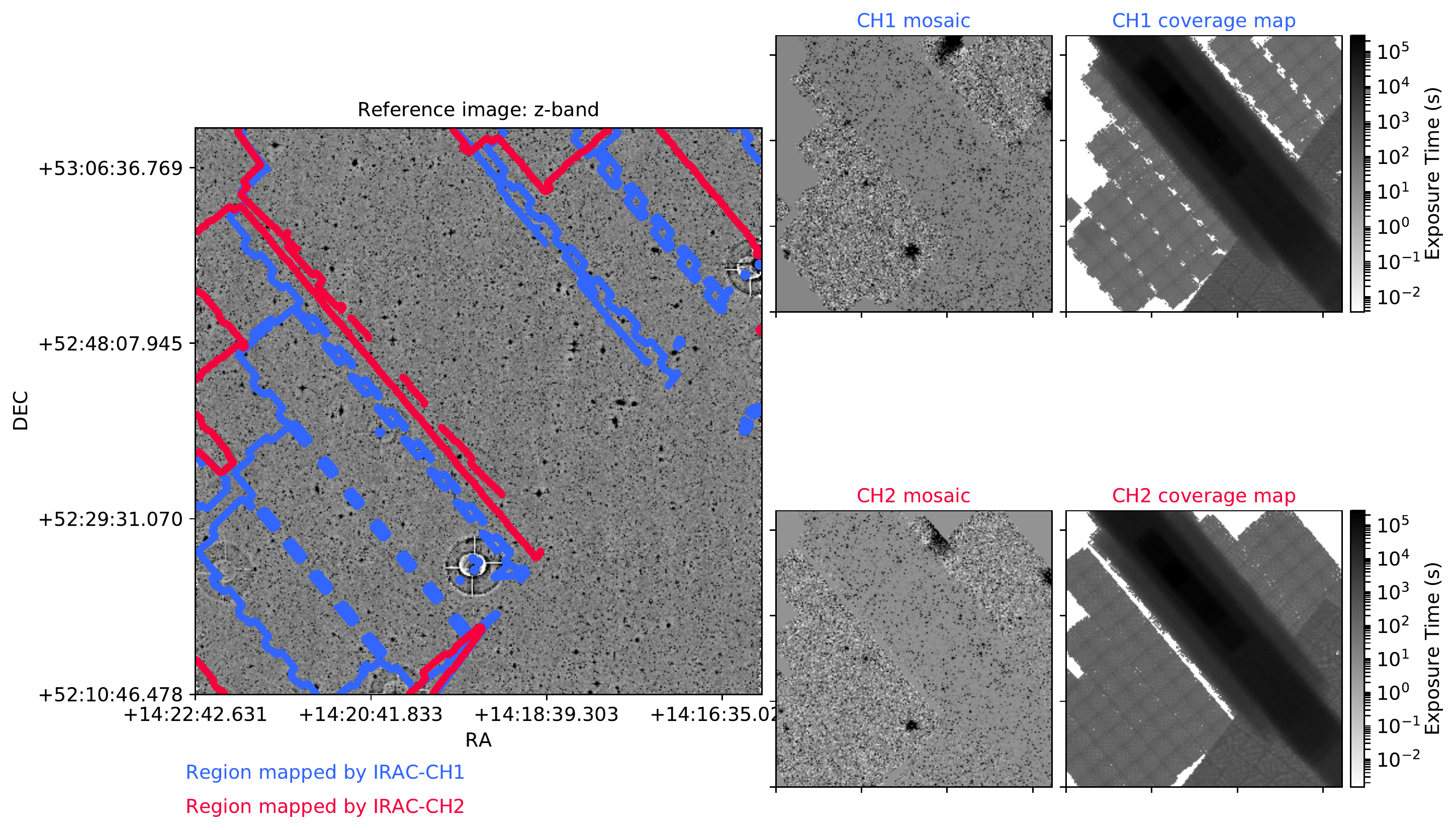}
	\caption{Same as in Figure~\ref{f:cfhtlsd1}, but for the CFHTLS-D3 field. For this field, we use a logarithmic scale for the coverage maps to show simultaneously the regions with the highest and lowest coverage.}
	\label{f:cfhtlsd3}
\end{figure*}

\begin{figure*}[ht]
	\centering
	\includegraphics[scale=0.5]{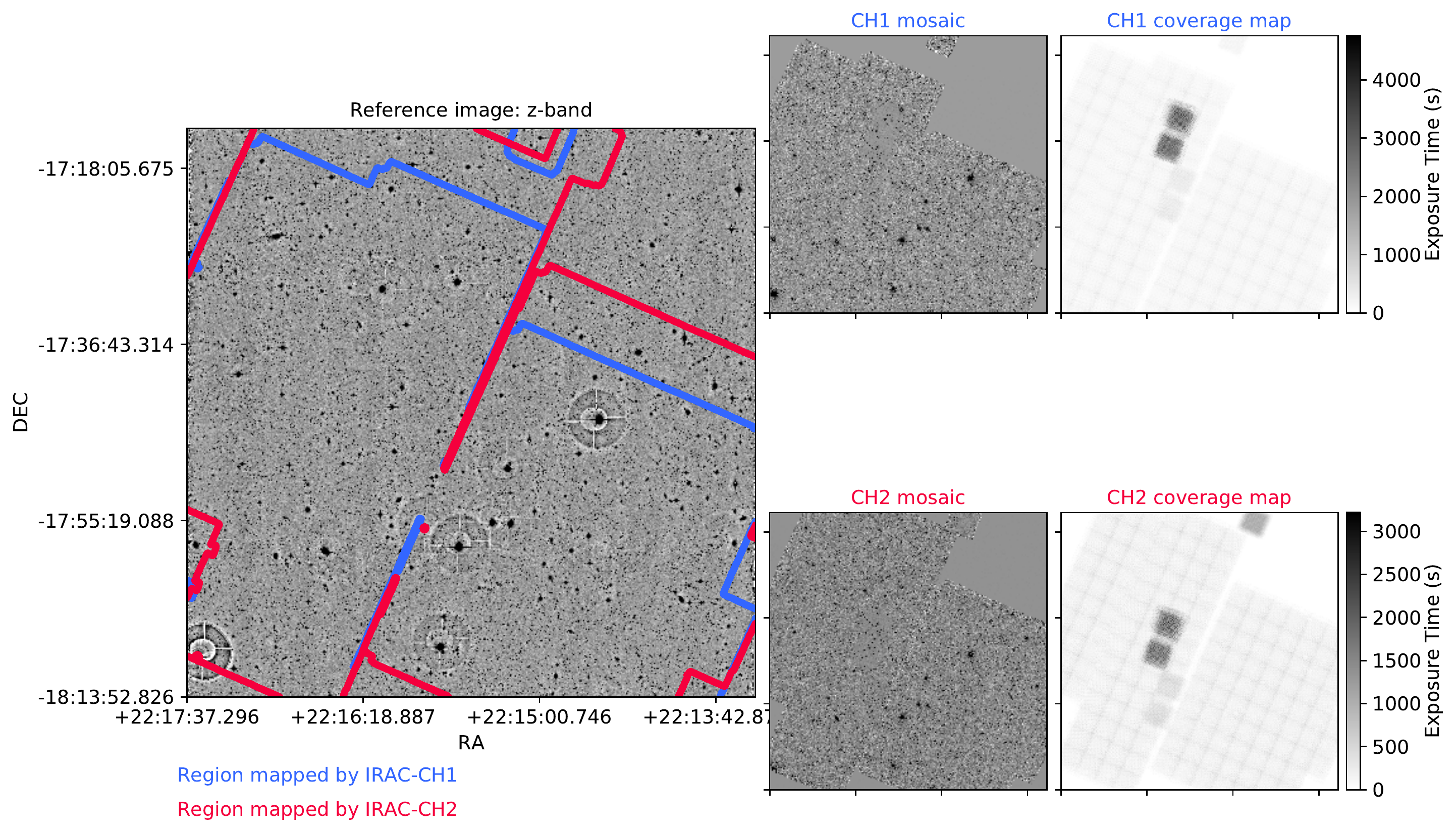}
	\caption{ Same as in Figure~\ref{f:cfhtlsd1}, but for the CFHTLS-D4 field. }
	\label{f:cfhtlsd4}
\end{figure*}

\begin{figure*}[ht]
	\centering
	\includegraphics[scale=0.5]{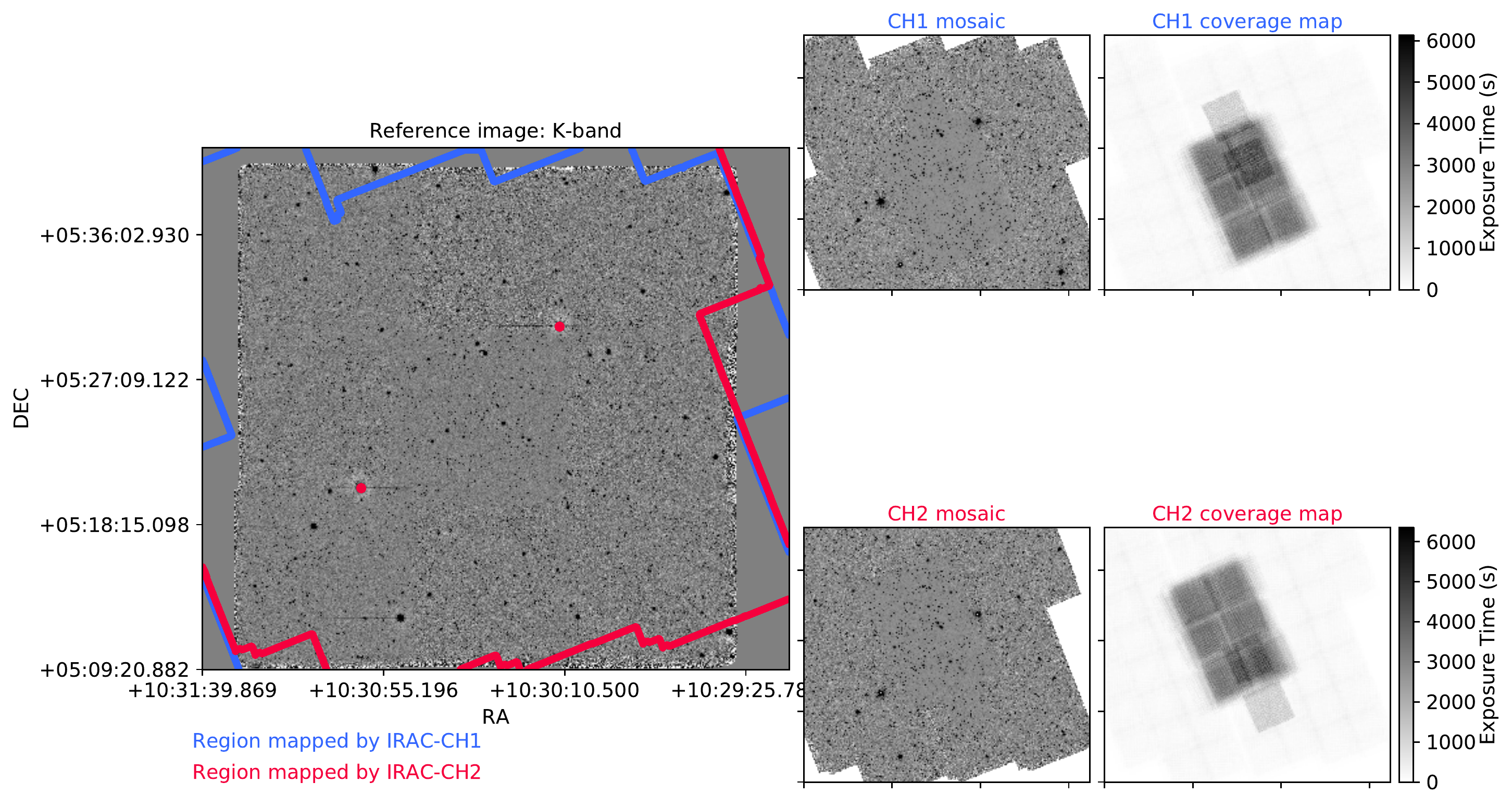}
	\caption{ Same as in Figure~\ref{f:cfhtlsd1}, but for the MUSYC1030 field.}
	\label{f:musyc1030}
\end{figure*}

\begin{figure*}[ht]
	\centering
	\includegraphics[scale=0.5]{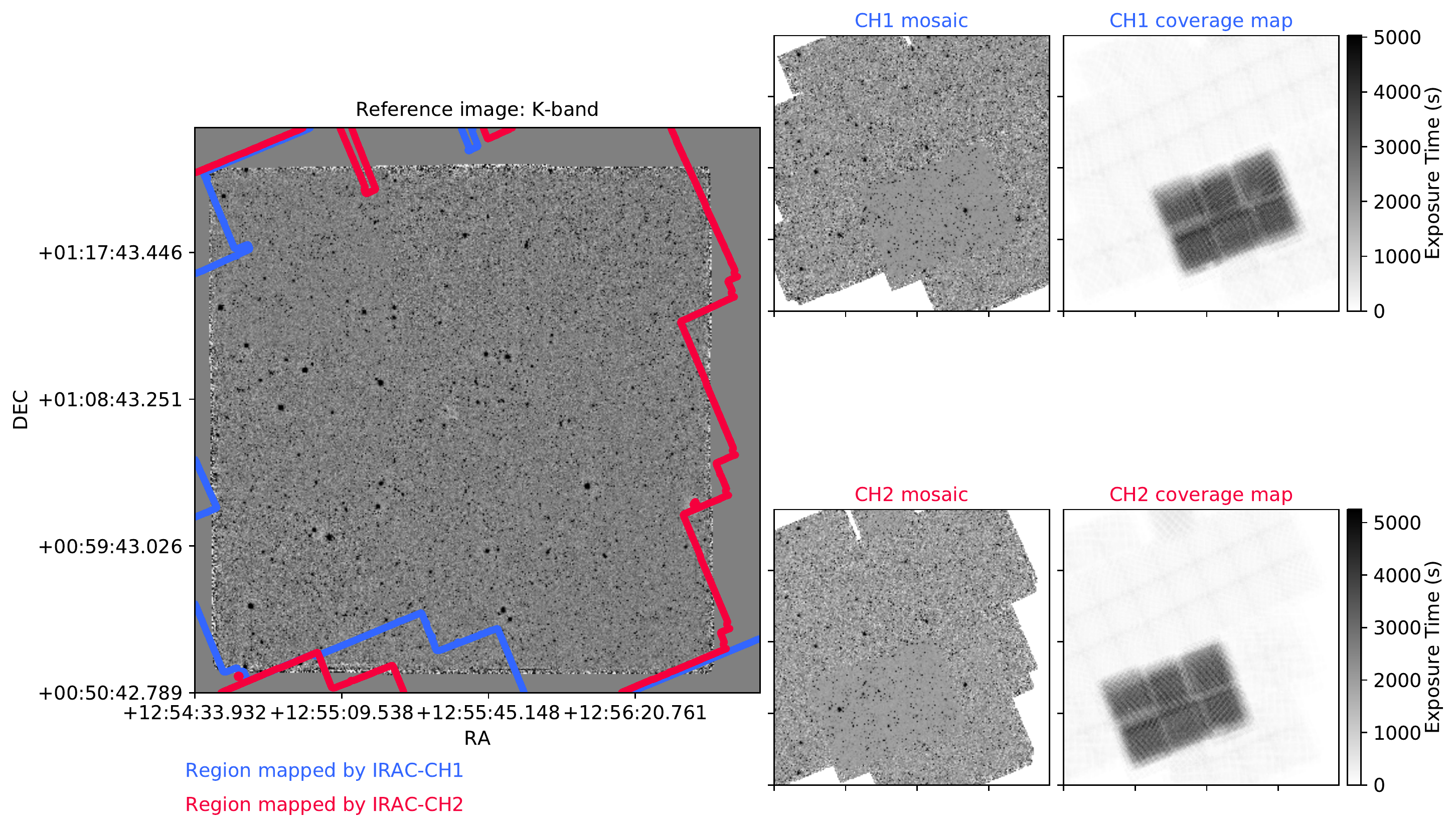}
	\caption{ Same as in Figure~\ref{f:cfhtlsd1}, but for the MUSYC1255 field. } 
	\label{f:musyc1255}
\end{figure*}

\begin{deluxetable*}{ccccccc}
\tablewidth{0pt}
\tablecaption{Characteristics of the IRAC observations
		\label{tab-appiracdata1}}
\tablehead{  \colhead{Filter} & \colhead{PIDs\tablenotemark{a}}  & \colhead{Max Exposure Time\tablenotemark{b}} & \colhead{Number of AORs}  &  \colhead{Pixel scale} & \colhead{ZP} \\
	    	  ($\mu$m) & & (h) & & (arcsec/pixel) & (AB) }
 \startdata 
% 	CFHTLS-D1: & &  & &\\
% 	3.6 &  21 & 262  & 0.558 & 19.03\\
% 	4.5 & 21 & 262  & 0.558 & 18.74\\
% 	\hline
% 	 CFHTLS-D3: &  & & &\\
% 	 3.6 & 787 & 456 & 0.558 & 20.08\\
% 	 4.5 & 787 & 456  & 0.558 & 20.13 \\
% 	 \hline
% 	 CFHTLS-D4: &  & & &\\
% 	 3.6 & 13 & 13  & 0.558 & 20.08 \\
% 	 4.5 & 13  & 13  & 0.558 & 20.08\\
 CFHTLS-D1: & & &  & &\\
3.6 & 181 & \textnormal{8.18} & 262  & 0.558 & 20.04\\
4.5 & 181 & \textnormal{7.31} & 262  & 0.558 & 20.04\\
 CFHTLS-D2\tablenotemark{c}: & & &  & &\\
3.6 & 20070, 61043,80057, 90042, 10159, 11016 &  &   & 0.6 & \\
4.5 & 20070, 61043,80057, 90042, 10159, 11016 &  &   & 0.6 & \\
\hline
CFHTLS-D3: & &  & & &\\
3.6 & 8, 41023, 61042, 80216, 10084 & \textnormal{29.3} & 456 & 0.558 & 20.04\\
4.5 & 8, 41023, 61042, 80216, 10084 & \textnormal{26.8} & 456  & 0.558 & 20.04 \\
\hline
CFHTLS-D4: & &  & & &\\
3.6 & 10084  & \textnormal{1.32} & 13  & 0.558 & 20.04 \\
4.5 & 10084  & \textnormal{0.89}  & 13  & 0.558 & 20.04\\
\hline
MUSYC-ECDF\tablenotemark{d}: & &  & & &\\
3.6 & 20708 & \textnormal{2.5} &   & 0.6 & 22.42\\
4.5 & 20708 & \textnormal{2.5} &   & 0.6 & 22.19\\
\hline
MUSYC1030: & &  & & &\\
3.6 & 3198, 10084 & \textnormal{1.70} & 11  & 0.267 & 20.04\\
4.5 & 3198, 10084 & \textnormal{1.76} & 11  & 0.267 & 20.04\\
\hline
MUSYC1255:&  &  & & &\\
3.6 & 30873, 10084 & \textnormal{1.40} & 13  & 0.269 & 20.04 \\
4.5 & 30873, 10084 & \textnormal{1.46} & 13  & 0.269 & 20.04\\
\hline
 \enddata
  \tablenotetext{a} {Spitzer Program Identification Numbers}
 \tablenotetext{b}{Maximum exposure time per position on sky per channel.}
 \tablenotetext{c}{The characteristics of this field are taken from \citealt{ashby2018}}
  \tablenotetext{d}{The characteristics of this field are taken from \citealt{damen2011}}
\end{deluxetable*}
\noindent In Figures \ref{f:cfhtlsd1}, \ref{f:cfhtlsd3}, \ref{f:cfhtlsd4}, \ref{f:musyc1030}, and \ref{f:musyc1255}, we show the reference image, the two IRAC mosaics at 3.6 $\mu  m$ and 4.5 $\mu  m$ and the corresponding coverage maps for each of the five fields.

%\begin{figure*}
%	\includegraphics[scale=0.7]{./figures_journal/all_fields.eps}
%	\caption{IRAC 3.6 $\mu$m mosaics for  CFHTLS-D1, CFHTLS-D3, CFHTLS-D4, MUSYC1030, and MUSYC1255.}
%	\label{f:fields}
%\end{figure*}

\section{Mosaic quality}\label{s:prop}
In this section, we describe the quality of the final mosaics  in both channels for  each of the five NMBS-II fields.

\begin{figure*}[h]
	\centering
	\includegraphics[scale=0.45]{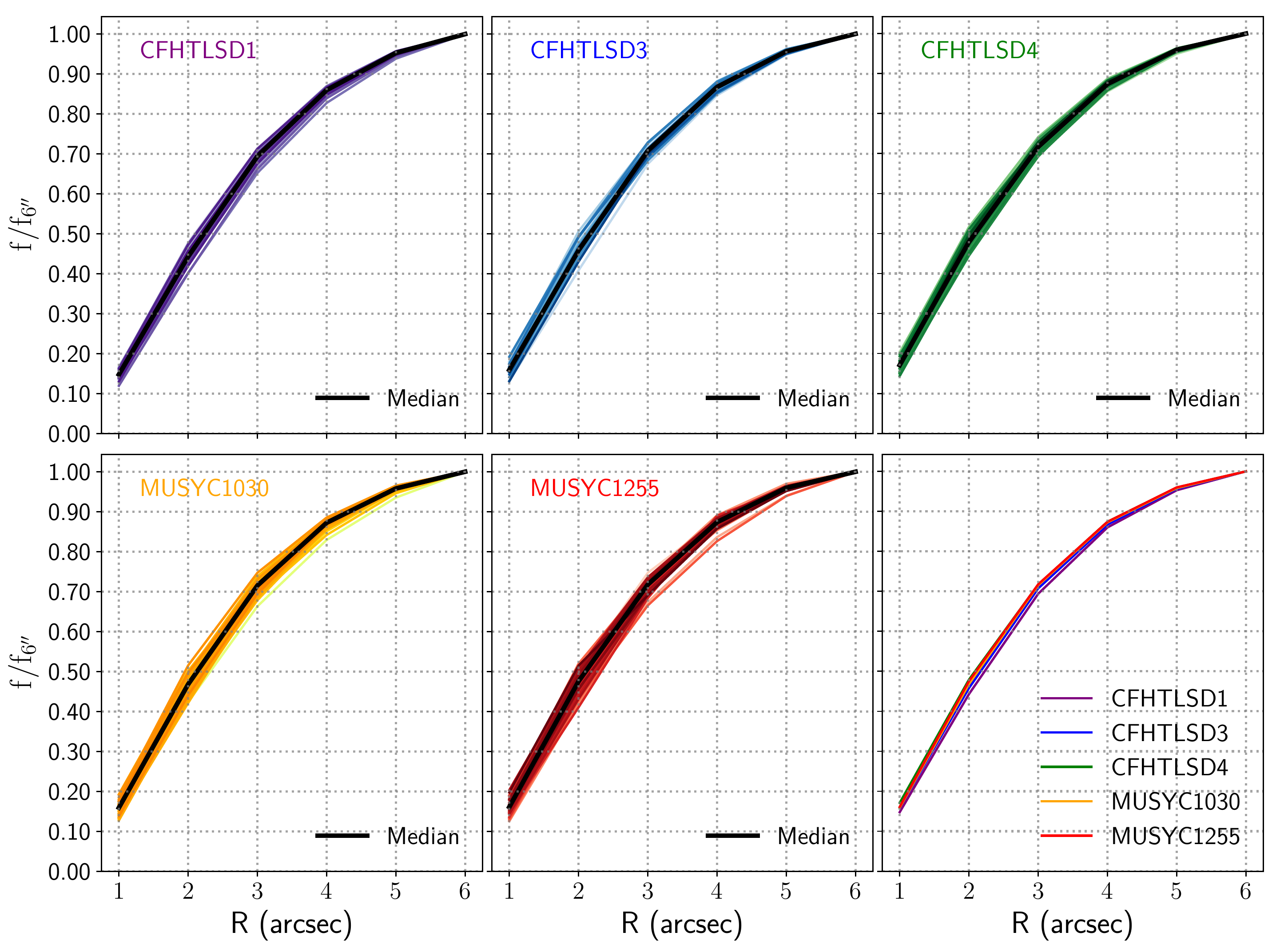}
	\caption{ First five panels: Growth curves for a sample of $\sim$ 50 bright unsaturated stars for all five fields at 3.6$\mathrm{\mu m}$ normalized to the flux of the star in an aperture of $\mathrm{6 \arcsec}$ diameter. The black curve in each panel represents the median of the distribution. The bottom-right panel shows the median growth curve of each field at  3.6$\mathrm{\mu m}$. }
	\label{f:psf_visual_1}
\end{figure*}

\begin{figure*}[h]
	\centering
	\includegraphics[scale=0.45]{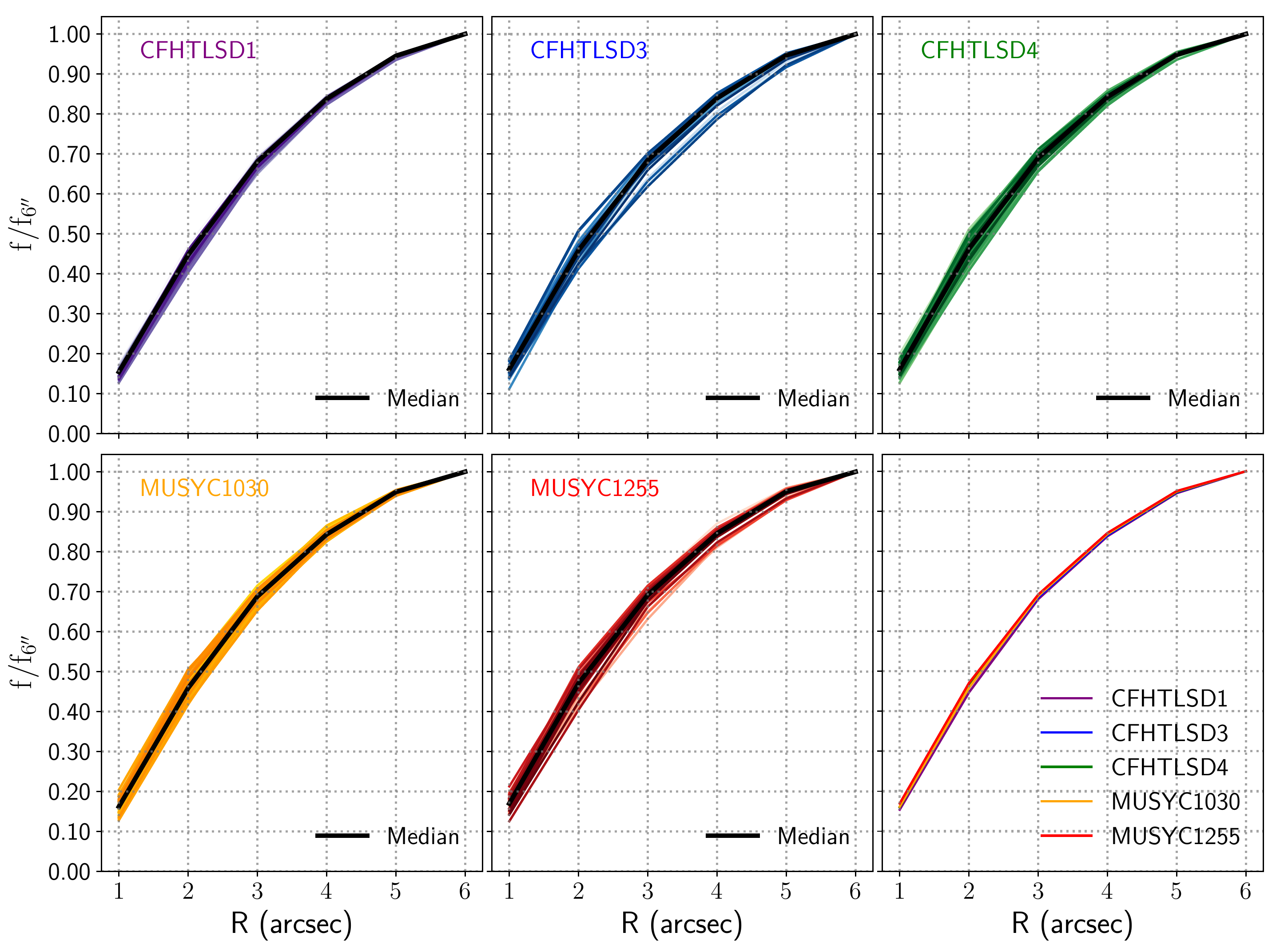}
	\caption{ Same as in Fig. ~\ref{f:psf_visual_1}, but at 4.5$\mathrm{\mu m}$. }
	\label{f:psf_visual_2}
\end{figure*}

\textnormal{To characterize the FWHM and the PSF of these mosaics we proceed in this way. First,  we define a sample  of $\sim$ 50 bright, isolated and unsaturated stars in each channel and each field. Then, for each star in the sample we measured a curve of growth,  i.e. the distribution of the light at different radii, normalized by the flux of the star in aperture of sufficiently large diameter ($\mathrm{6 \arcsec}$). For each field,  we consider the median of this growth curves in each channel (black curves in Fig.~\ref{f:psf_visual_1} and Fig.~\ref{f:psf_visual_2}). Outliers, such as saturated stars, were then determined based on the shape of their light profile compared with the median curve of growth, and rejected from the sample.  We use the median growth curve in each field to derive the median half-light and 75\%-light radii, i.e.,  the radii that contain 50\% and 75\% of the light,  respectively. These values are reported in Table  ~\ref{t:psf}.  The same sample of stars is also used to derive the FWHMs of the IRAC images. The derived FWHM are listed in Table ~\ref{t:psf}. We stress that the IRAC PSF is complicated and that spatial variations across the field of view are to be expected given we mosaiced IRAC data taken from several independent programs. However, the state-of-the-art approach to extract IRAC photometry from wide and deep surveys consists in the adoption of software that accounts for spatially varying PSF across the field of view, such as MOPHONGO \citep{labbe2006}, TFIT \citep{laidler2007}, and T-PHOT \citep{merlin2015}. } 

\begin{figure*}[ht]
	\centering
	\includegraphics[scale=0.4]{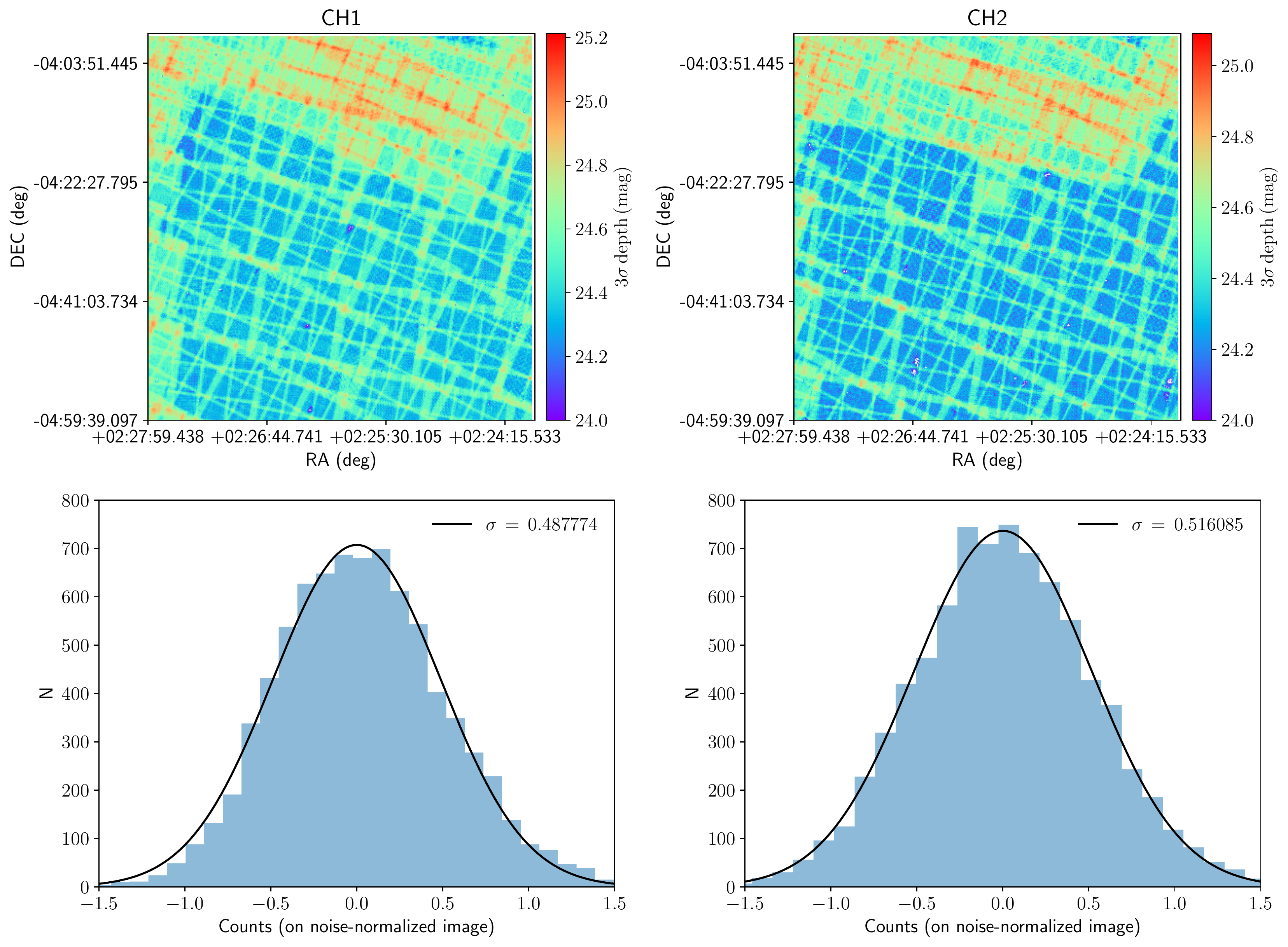}
	\caption{CFHTLSD1: Top row: 3$\sigma$ magnitude depths of $\mathrm{3.6\, \mu m}$ and $\mathrm{4.5\, \mu m}$ mosaics. Bottom row: Histograms of enclosed fluxes measured on $\sim$ 8000 empty apertures in $\mathrm{3.6\, \mu m}$ and $\mathrm{4.5\, \mu m}$ on the noise-normalized images. Best-fit Gaussians are also shown. }
	\label{f:empty_cfhtlsd1}
\end{figure*}

\begin{figure*}[ht]
	\centering
	\includegraphics[scale=0.4]{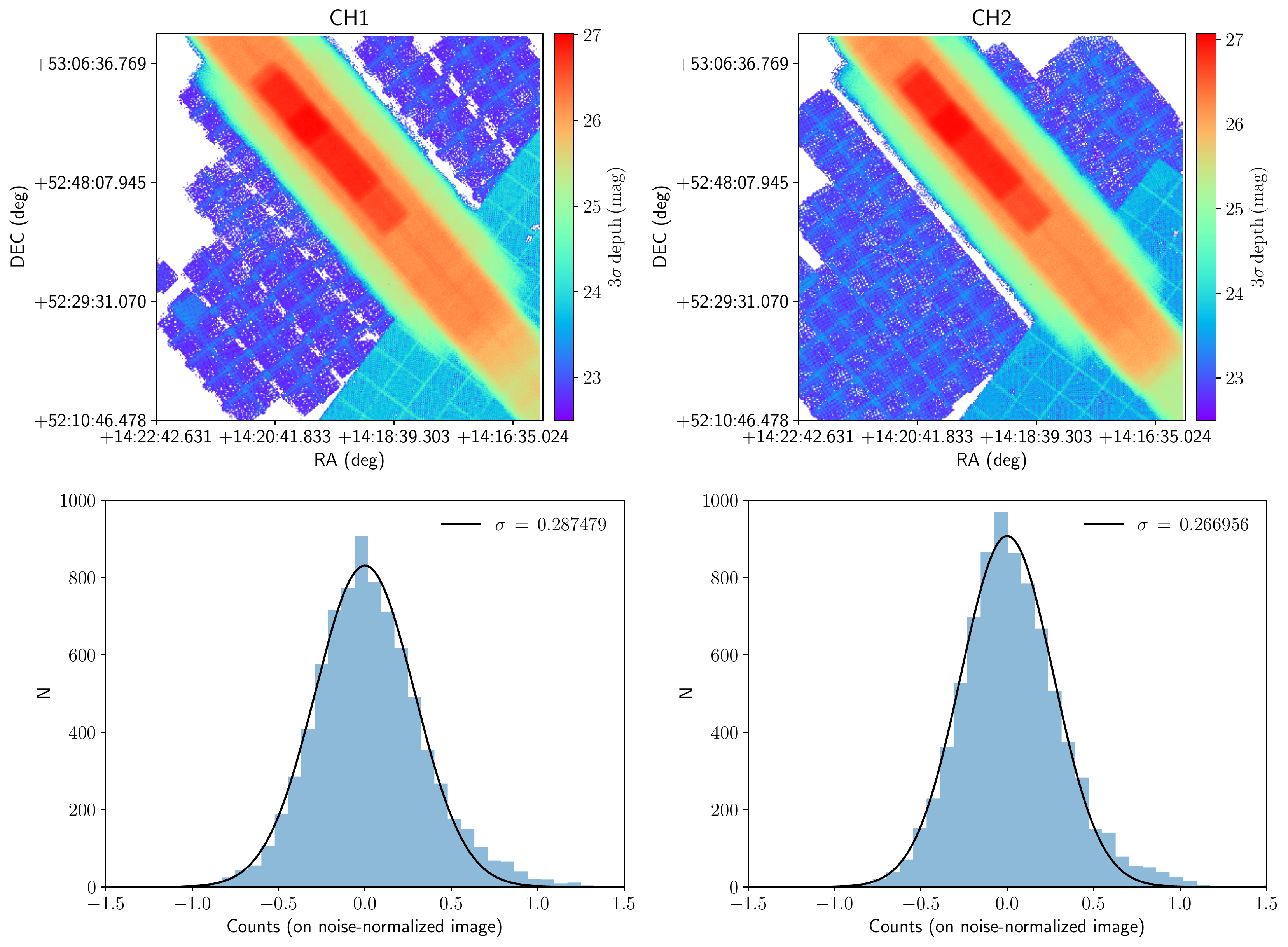}
	\caption{ Same as in Figure~\ref{f:empty_cfhtlsd1}, but for the CFHTLS-D3 field.}
	\label{f:empty_cfhtlsd3}
\end{figure*}

\begin{figure*}[ht]
	\centering
	\includegraphics[scale=0.4]{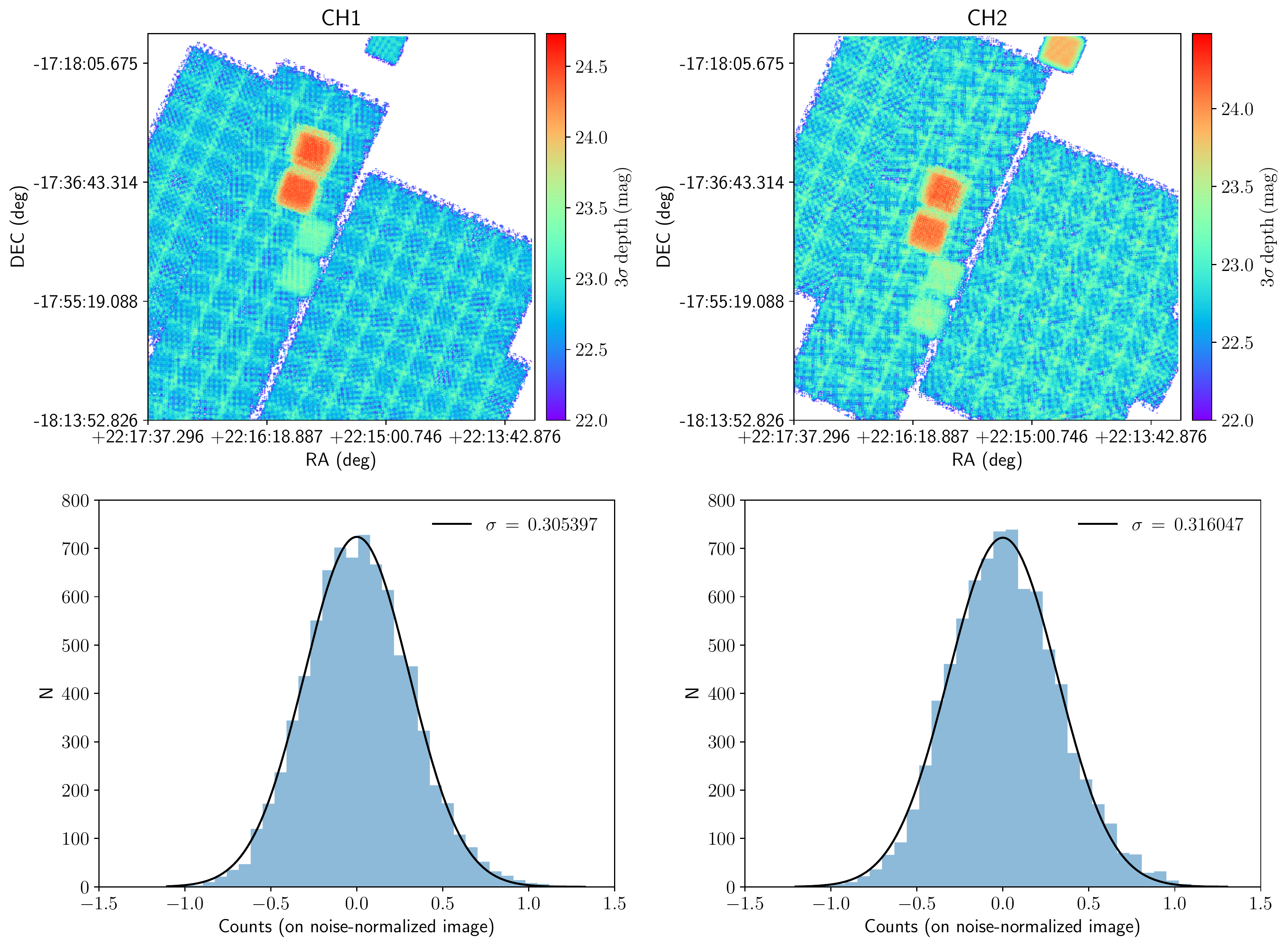}
	\caption{ Same as in Figure~\ref{f:empty_cfhtlsd1}, but for the CFHTLS-D4 field.}
	\label{f:empty_cfhtlsd4}
\end{figure*}

\begin{figure*}[ht]
	\centering
	\includegraphics[scale=0.4]{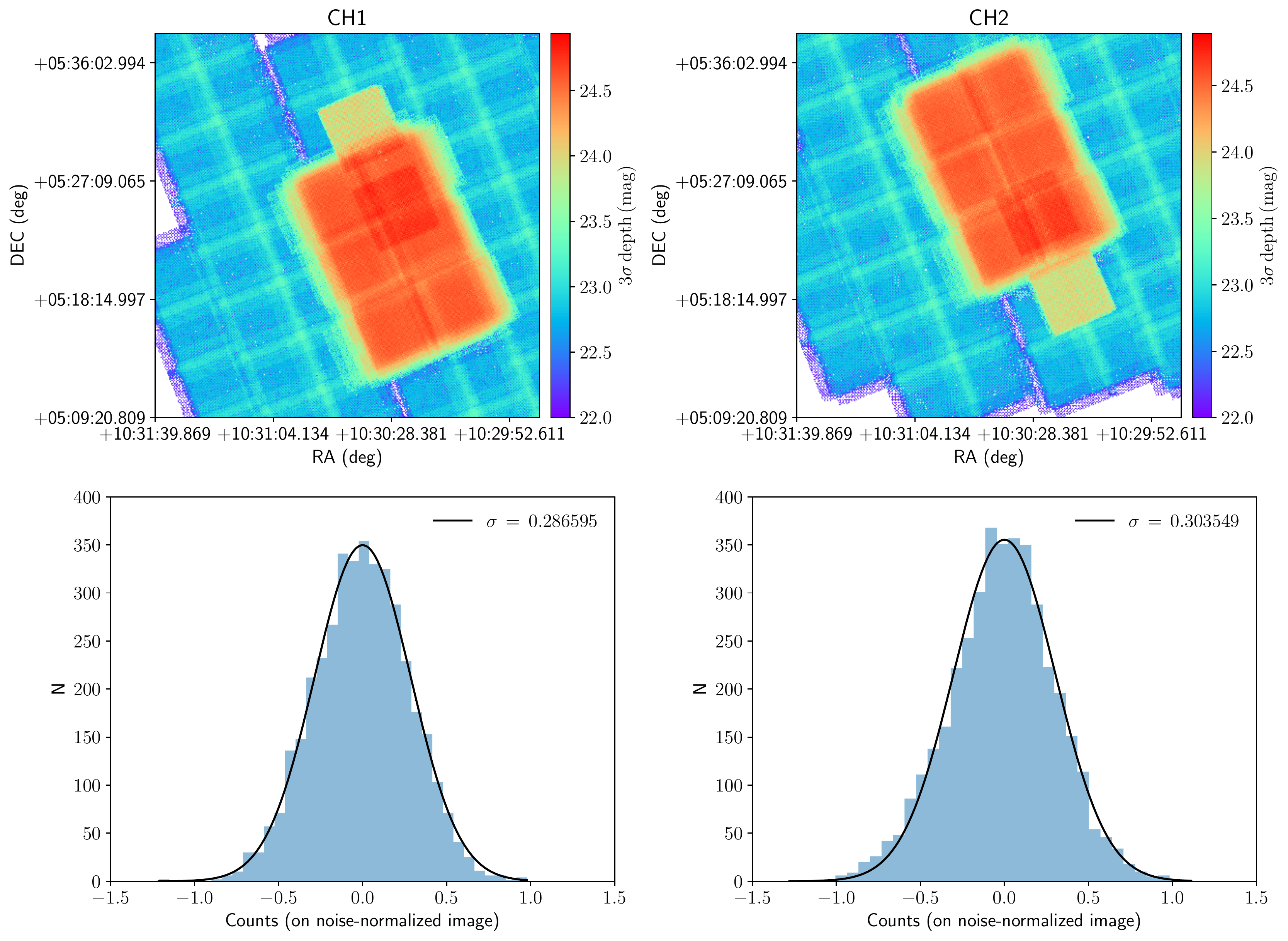}
	\caption{MUSYC1030: Top row: 3$\sigma$ magnitude depths of $\mathrm{3.6\, \mu m}$ and $\mathrm{4.5\, \mu m}$ mosaics. Bottom row: Histograms of enclosed fluxes measured on $\sim$ 4000 empty apertures in $\mathrm{3.6\, \mu m}$ and $\mathrm{4.5\, \mu m}$ on the noise-normalized images. Best-fit Gaussians are also shown. }
	\label{f:empty_musyc1030}
\end{figure*}

\begin{figure*}[ht]
	\centering
	\includegraphics[scale=0.4]{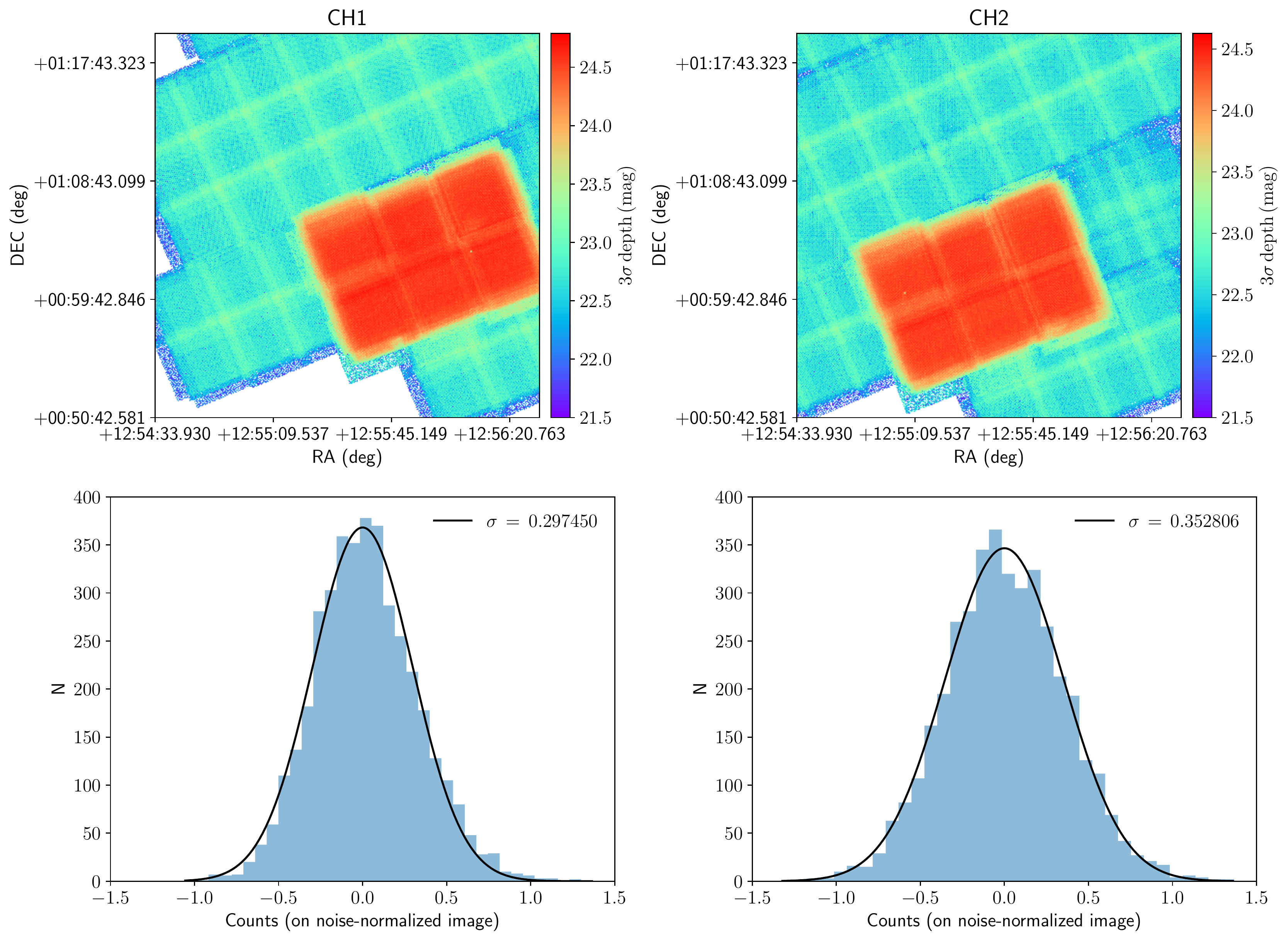}
	\caption{Same as in Figure~\ref{f:empty_musyc1030}, but for the MUSYC1255 field. }
	\label{f:empty_musyc1255}
\end{figure*}

%FWHMs of the IRAC images were derived by taking the median of the FWHM for a set of $\sim$30 bright, isolated stars in each field. The derived FWHM are listed in Table ~\ref{t:ap}.
In Table ~\ref{t:ap}, we list the  3$\sigma$ magnitude depth for all mosaics.  For comparison, we also add, where available, the FWHMs and  3$\sigma$ magnitude depths of the other two fields taken from the literature papers mentioned in Sect.~\ref{s:obs}. \\
%These values are determined sby using the ``empty aperture'' method, following the same approach as for the deep NIR MUSYC data \citep{quadri2007}. 
\noindent We use the ``empty aperture'' method to empirically determine the noise properties of our IRAC mosaics, following the same approach as for the deep NIR MUSYC data \citep{quadri2007}.  
Briefly, we randomly place a large number of apertures on the noise-normalized images (obtained by multiplying the images by the square root of the coverage maps), reject all apertures falling on sources, and measure the flux in the remaining ones. We chose an aperture size of $3\arcsec$ and measure the flux in $\sim$ 8000 apertures for CFHTLSD fields and  $\sim$ 4000 in the MUSYC fields.  A histogram of the measured fluxes is constructed for each field  in both $\mathrm{3.6\, \mu m}$ and $\mathrm{4.5\, \mu m}$, and shown in Figures~\ref{f:empty_cfhtlsd1}, ~\ref{f:empty_cfhtlsd3}, ~\ref{f:empty_cfhtlsd4}, ~\ref{f:empty_musyc1030} and ~\ref{f:empty_musyc1255} (bottom panels). 
These histograms are all well fitted by a Gaussian.  The top panels of Figures ~\ref{f:empty_cfhtlsd1}, ~\ref{f:empty_cfhtlsd3}, ~\ref{f:empty_cfhtlsd4}, ~\ref{f:empty_musyc1030} and ~\ref{f:empty_musyc1255} show  the maps of  $\mathrm{3\sigma}$ magnitude depths derived using the empty aperture method according to the Equation:

\begin{equation}
\mathrm{depth(3\sigma)[AB] = -2.5 \log{(\frac{3\sigma_{best-fit}}{\sqrt{\mathrm{COVERAGE\, MAP}}})} + ZP,}
\end{equation}

where ZP is listed in Table~\ref{tab-appiracdata1} for each field.
Table ~\ref{t:ap} lists the 15th, 50th, and 75th percentiles of the 3$\sigma$ magnitude depths in both $\mathrm{3.6\, \mu m}$ and $\mathrm{4.5\, \mu m}$ in an aperture of D=3$\arcsec$ as derived with the empty aperture method. The median value of the 3$\sigma$ depths for $\mathrm{3.6\, \mu m}$ varies from 22.83 mag (in CFHTLSD4 ) to 24.52 (in CFHTLSD1), while for $\mathrm{4.5\, \mu m}$ it varies from 22.85 mag (in CFHTLSD4) to 24.44 (in CFHTLSD1 ).
As expected, regions with higher exposure times (see Figures~\ref{f:cfhtlsd1}, ~\ref{f:cfhtlsd3}, ~\ref{f:cfhtlsd4},~\ref{f:musyc1030}, and \ref{f:musyc1255})  also have fainter $3\sigma$ magnitude depths. As shown in Figure~\ref{f:empty_cfhtlsd3}, using both the images observed in this program and the data collected by the AEGIS survey, we present results in an extremely deep mosaic in some of the regions of CFHTLSD3.  \\  

\begin{deluxetable*}{cccc}
	\tablewidth{0pt}
	\tablecaption{Characteristics of the IRAC observations
		\label{t:psf}}
	\tablehead{\colhead{Filter} & \colhead{FWHM} & \colhead{50\% light radius}  & \colhead{75\% light radius}  \\
		($\mu$m)      & (\arcsec)      & ( \arcsec)   & ( \arcsec) }
	\startdata 
	CFHTLS-D1:& & &   \\
	3.6 &  2.08 & 2.20 & 3.28  \\
	4.5 &  1.95 & 2.20 & 3.39  \\
	\hline
	CFHTLS-D2\tablenotemark{a}:& & &  \\
	3.6 &  $\mathrm{\sim}$ 2.00 &  &   \\
	4.5 &  $\mathrm{\sim}$ 2.00 &  &    \\
	\hline
	CFHTLS-D3:& & & \\
	3.6 & 2.00 & 2.15 & 3.22\\
	4.5 & 1.86 & 2.16 & 3.38  \\
	\hline
	CFHTLS-D4:& & & \\
	3.6 & 1.86 & 2.08 & 3.17  \\
	4.5 & 1.75 & 2.14 & 3.35 \\
	\hline
	MUSYC-ECDF\tablenotemark{b}:& & &\\
	3.6 & 1.97 & &\\
	4.5 & 1.93 & & \\
	\hline
	MUSYC1030:& & &\\
	3.6 & 2.17 & 2.11 & 3.18  \\
	4.5 & 2.01 & 2.16 & 3.35  \\
	\hline
	MUSYC1255:& & &\\
	3.6 & 2.02 & 2.09 & 3.17 \\
	4.5 & 1.98 & 2.11 & 3.34  \\
	\hline
	\enddata
	\tablenotetext{a}{The FWHM of this field  are taken from \citealt{ashby2018}.}
	\tablenotetext{b}{The FWHM of this field are taken from \citealt{damen2011}. }	
\end{deluxetable*}

\begin{deluxetable*}{cccc}
	\tablewidth{0pt}
	\tablecaption{Depths of the IRAC observations
		\label{t:ap}}
	\tablehead{\colhead{Filter} & \colhead{15th percentile of 3$\sigma$ depth\tablenotemark{a}}  & \colhead{Median of 3$\sigma$ depth \tablenotemark{a}} & \colhead{75th percentile of 3$\sigma$ depth\tablenotemark{a}} \\
		($\mu$m)          & ( AB)   & ( AB)  & ( AB)}
	\startdata 
	CFHTLS-D1:& & &  \\
	3.6 &   24.33 & 24.52 & 24.66  \\
	4.5 &   24.24 & 24.44 &  24.58  \\
	\hline
	CFHTLS-D2\tablenotemark{b}:&  & &  \\
	3.6 &    & 25.31 &   \\
	4.5 &    & 25.31  &   \\
	\hline
	CFHTLS-D3:& & &\\
	3.6 & 22.79 & 23.76 &  25.54\\
	4.5 &  22.88 & 23.45 & 25.29  \\
	\hline
	CFHTLS-D4:&  & & \\
	3.6 &  22.53 & 22.83 & 23.00 \\
	4.5 &  22.51 & 22.85 &  23.03 \\
	\hline
	MUSYC-ECDF\tablenotemark{c}:& & &\\
	3.6 & 24.21  & 24.41 & 24.55 \\
	4.5 &  24.05 & 24.24 &  24.37 \\
	\hline
	MUSYC1030:&  & &\\
	3.6 &  22.63 & 22.92 & 23.20 \\
	4.5 &  22.63 & 22.97 &  23.24 \\
	\hline
	MUSYC1255:&  & &\\
	3.6 &  22.60 & 22.87 & 23.10  \\
	4.5 &  22.44 & 22.75 & 23.00   \\
	\hline
	\enddata
	\tablenotetext{a}{Using a 3$\arcsec$ circular aperture diameter.}
	\tablenotetext{b}{The 3$\sigma$ depths of this field  are taken from \citealt{ashby2018}.}
	\tablenotetext{c}{The  3$\sigma$ depths of this field are taken from \citealt{damen2011}. }
	
\end{deluxetable*}

\section{Public Data release}\label{s:release}
The data release consists of reduced images of all IRAC observations in three CFHTLSD fields, namely CFHTLS-D1, CFHTLS-D3, CFHTLS-D4,  and two MUSYC fields, MUSYC1030, and MUSYC1255. The images are available on the website: \url{http://cosmos.phy.tufts.edu/~danilo/NMBS2irac}.
The data release contains the following:
\begin{itemize}
\item Mosaic science images in both 3.6 $\mu$m and 4.5 $\mu$m. The images are W-registered to the reference image. This reference image is either the z-band stack image of the 25\% best seeing images from the CFHTLSDeep for CFHTLS-D1, CFHTLS-D3, CFHTLS-D4, or the K-band images obtained at the Blanco 4-m telescope at CTIO for MUSYC1030 and MUSYC1255.
\item Coverage maps containing exposure times (seconds) in both 3.6 $\mu$m and 4.5 $\mu$m. 
\item Maps with 3$\sigma$ magnitude depth (as derived in Section~\ref{s:prop}) at the same WCS and pixel scales of the scientific images.
%\item  Reduced images of all individual 353 AORs, drizzled onto the same grid, which may be useful to study the  reliability or variability of sources.
\end{itemize}

\section*{Acknowledgment}
The reduction and public release of the Spitzer-IRAC data in the NEWFIRM Medium-Band Survey II was supported by the National Aeronautics and Space Administration (NASA) under award number NNX16AN49G issued through the NNH15ZDA001N Astrophysics Data Analysis Program (ADAP). The Cosmic Dawn Center is funded by the Danish National Research Foundation. We thank the anonymous referee for his/her useful comments.

\bibliographystyle{aa}

\bibliography{bibliography}

\end{document}